\documentclass[print]{aa}

\usepackage{graphicx}

\usepackage{natbib}
\usepackage{txfonts}

\begin{document}

\title {Detailed Chemical Evolution of Carina and Sagittarius
 Dwarf Spheroidal Galaxies}
\titlerunning{Chemical Evolution of Carina and Sagittarius}

\author{Gustavo A. Lanfranchi\inst{1}
\and    Francesca  Matteucci\inst{2, 3}
\and    Gabriele Cescutti\inst{2}}

\institute{IAG-USP, R. do Mat\~ao 1226, Cidade Universit\'aria, 
05508-900 S\~ao Paulo, SP, Brazil
\and Dipartimento di Astronomia-Universit\'a di Trieste, Via G. B. 
Tiepolo 11, 34131 Trieste, Italy
\and  I.N.A.F. Osservatorio Astronomico di Trieste, via G.B. Tiepolo 
11, I-34131}

\date{Received xxxx/ Accepted xxxx}

\abstract{}
{
In order to verify the effects of the
most recent data on the evolution of Carina and Sagittarius 
Dwarf Spheroidal Galaxies (dSph) and to set 
tight constraints on the main parameters of chemical evolution models,
we study in detail the chemical evolution of these galaxies
through comparisons between the new data and the predictions of 
a model, already tested to reproduce 
the main observational constraints 
in dSphs. }
{
Several abundance ratios, such as [$\alpha$/Fe],
[Ba/Fe] and [Eu/Fe], and the metallicity 
distribution of stars are compared to the predictions 
of our models adopting the observationally derived star 
formation histories in these
galaxies.}
{
These new comparisons confirm our previously
suggested scenario for the evolution of these galaxies, 
and allow us to better fix the star formation and wind parameters. 
In particular, for Carina our predictions indicates that 
the best efficiency 
of star formation is $\nu = 0.15 \;Gyr^{-1}$, that the 
best 
wind efficiency parameter is  $w_i$ = 5 (the wind rate is five times 
stronger than the star formation rate), and that the star formation history,
which produces the best fit to the observed metallicity distribution of 
stars is characterized by several episodes of activity.
In the case of Sagittarius our results suggest that 
$\nu=3\; Gyr^{-1}$ and $w_i=9$, again in agreement with our 
previous work. 
Finally, we show new predictions for [N/Fe] and [C/Fe] ratios for the 
two galaxies suggesting a scenario for Sagittarius very similar 
to the one of the solar vicinity in the Milky Way, 
except for a slight decrease of [N/Fe] ratio at high metallicities
due to the galactic wind.
For Carina we predict a larger [N/Fe] ratio at low metallicities,
reflecting the lower star formation efficiency of this galaxy
relative to Sagittarius and the Milky Way.}
{} 

\keywords {stars: abundance --  galaxies: abundances 
-- galaxies: Local Group -- galaxies: evolution -- galaxies: dwarf -- }

\maketitle

\section{Introduction}
In the last few years a large amount of observing time was devoted to the 
analysis of Dwarf Spheroidal Galaxies of the Local Group.
The study of these galaxies is favored by their relative 
proximity, which  enables one to obtain several observational 
constraints with high accuracy. The determination of the 
star formation histories inferred through
observed colour-magnitude diagrams (Smecker-Hane et al. 1996;
Hurley-Keller, Mateo, Nemec 1998; Hernandez, Gilmore,
Valls-Gabaud 2000; Dolphin 2002; Rizzi et al. 2003), the gas mass, 
the total mass, the photometric 
properties (Mateo 1998 and references therein), the elemental 
abundances and  abundance ratios (Bonifacio et al. 2000; Shetrone, 
Cot\'e, Sargent 2001; Shetrone et al. 2003; Tolstoy et al. 2003,
Bonifacio et al. 2004; 
Venn et al. 2004; Sadakane et al. 2004; Geisler et al. 2005;
Monaco et al. 2005) and the metallicity distribution of stars 
(Koch et al. 2005) can help us in  
understanding the formation and evolution of the dSph galaxies.

From the observations emerged a scenario in which the dSphs are 
characterized by low star formation (SF) efficiencies,
with complex star formation histories (SFH), where 
some galaxies exhibit old metal-poor, young metal-rich, 
intermediate, or even a mixed populations of stars,
and by a central region almost totally depleted of neutral gas. 
In order to 
reproduce this scenario and fit some of the observational
constraints, a few chemical evolution models were proposed in 
the past years
(e.g. Ikuta $\&$ Arimoto 2002; Carigi, Hernandez, Gilmore 2002;
 Lanfranchi $\&$ Matteucci 2003, 2004 - 
LM03, LM04). In particular, LM03 and LM04 succeeded in reproducing 
several abundance ratios (such as [Mg/Fe], [O/Fe], [Ca/Fe], [Si/Fe]),
the total mass and the gas content of six dSphs of the Local Group
(Carina, Draco, Sagittarius, Sculptor, Sextan, and Ursa Minor).
Besides that, they predicted the stellar metallicity distributions 
of these galaxies,  but at that time no such observations were available.
In their
scenario, the evolution of these dSph galaxies was controlled
mainly by the low star formation (SF) efficiency ($\nu = 0.01-0.5 
\;Gyr^{-1}$)
and by the high wind efficiency (6 to 13 times the star formation 
rate). These two
parameters were suggested to be the main responsibles for the 
observed abundance ratio patterns and for the shape and the
location of the peak of the stellar metallicity distributions
(see LM04 for more details).

Recently, many more new data concerning either abundance ratios or 
stellar metallicity distributions have become available for Carina 
and Sagittarius galaxies. 
In particular, 
Koch et al. (2005) observed almost 500 stars
in Carina and inferred the [Fe/H] abundance through the calcium triplet
lines. Now that these data are available we therefore intend to obtain
the best fit to stellar metallicity 
distribution of Carina and, at the same time, reproduce the observed 
abundance 
ratios, in order to better constrain the range of values suggested 
by LM04 for the efficiency of SF and the wind efficiency. 

We have the same goal in the case of Sagittarius. 
In fact, Bonifacio et al. (2004) and Monaco et al. (2005) recently 
published 
data on abundance ratios of several chemical species, not available 
at the times of our previous papers. They found very
low values of [$\alpha$/Fe] at moderate high metallicities 
(-1.0 $<$ [Fe/H] $<$ 0.0 dex)
and suggested that these values of [$\alpha$/Fe] indicate a low
SF rate, in contrast to what was proposed by the models of LM03, 
which require a SF efficiency between $\nu = 1 - 5\; Gyr^{-1}$ 
in order to reproduce the [$\alpha$/Fe] observed by
Smecker-Hane $\&$ Mc. William (1999) and Bonifacio et al. (2000).
Here we compare the same model for Sagittarius as described 
in LM03 and LM04
with the newest data of Bonifacio et al. (2004) and Monaco et al. 
(2005) with the aim of testing if the range of values proposed by
LM 03 and LM04 still holds or if they need a revision in the light of
the new data, as claimed by Monaco et al. (2005).

The paper is organized as follows: in Sect. 2 we present
the observational data concerning Carina and Sagittarius dSph 
galaxies, in Sect. 3 we summarize the main characteristics of 
the adopted chemical evolution models, such as
the star formation and the nucleosynthesis prescriptions,
in Sect. 4 the predictions of our models are 
compared to the recent observational data and the results discussed, 
and finally in Sect. 5 we draw some conclusions. All elemental
abundances are normalized to the solar values
([X/H] =  log(X/H) - log(X/H)$_{\odot}$)  as measured by 
Grevesse $\&$ Sauval (1998).

\section{Data Sample}

We gathered, together with the previous data of Smecker-Hane \& 
Mc. William(1999), the most recent data for
Sagittarius and Carina from the literature, including 
several abundance ratios in Sagittarius (Bonifacio et al. 2000, 
2004; Monaco et al. 2005) and abundance ratios (Shetrone, 
Cot\'e $\&$ Sargent 2001; Shetrone et al. 2003; Venn et al. 2004)
and the stellar metallicity distribution in Carina
(Koch et al. 2005). Bonifacio et al. (2000, 
2004) and Monaco et al. (2005) derived the abundances
of several chemical elements in 27 stars of Sagittarius in the metallicity
range
-1.52 $<$ [Fe/H] $<$ -0.17. Among these stars only two (from Bonifacio
et al. 2000) were used in the previous comparisons with the models in
LM03 and LM04. The remaining stars will now be put together with the 
previous two,
in order to see whether the predictions of our models can still reproduce 
the new data. For Carina, we considered the recent metallicity 
distribution of Koch  et al. (2005) together with the updated data 
of Venn et al. (2004) for several abundance ratios.

\begin{table*}
\begin{center}\scriptsize  
\caption[]{Models for dSph galaxies. $M_{tot}^{initial}$ 
is the baryonic initial mass of the galaxy, $\nu$ is the star-formation 
efficiency, $w_i$ is the wind efficiency,  $n$, $t$ and $d$ 
are the number, time of occurrence and duration of the SF 
episodes, respectively, and $t_{GW}$ the time of the occurrence of the 
Galactic Wind.}
\begin{tabular}{lcccccccc}  
\hline\hline\noalign{\smallskip}  
galaxy &$M_{tot}^{initial} (M_{\odot})$ &$\nu(Gyr^{-1})$ &$w_i$
&n &t($Gyr$) &d($Gyr$) &$t_{GW}(Gyr)$ &$IMF$\\    

\noalign{\smallskip}  
\hline
Carina &$5*10^{8}$ &0.15  &5 &4 &0/2/7/9 &2/2/2/2 &0.5 &Salpeter\\
Sagittarius &$5*10^{8}$ &1.0-5.0 &9-13 &1 &0 &13&0.05-0.15 &Salpeter\\
\hline\hline
\end{tabular}
\end{center}
\end{table*} 

\section{Models} 

In order to verify the validity of the models for Carina and Sagittarius
relative to the new data, we use the same prescriptions 
as described in LM03 and LM04. 
These models already reproduce very well several [$\alpha$/Fe] ratios, the 
[Ba/Fe] and [Eu/Fe] ratios (Lanfranchi, Matteucci $\&$
Cescutti 2005 - LMC05) as well as 
the present total mass and gas mass observed in these systems.
In the adopted scenario, the dSphs 
form through a continuous and fast infall of pristine gas until 
a mass of $\sim 10^8 M_{\odot}$ is accumulated. The SF
is characterized by
one long episode in the case of Sagittarius and by several episodes for
Carina, with low (in the 
case of Carina) to intermediate-high (for Sagittarius) values of the SF 
efficiency
and high galactic wind efficiency in both cases. In fact, the 
galactic winds
play a crucial role in the evolution of these galaxies. They
develop when the thermal 
energy of the gas equates its binding energy (see for example Matteucci
$\&$
Tornamb\'e 1987). This quantity is strongly influenced by 
assumptions concerning the presence and distribution 
of dark matter (Matteucci 1992). A diffuse ($R_e/R_d$=0.1, 
where $R_e$ is the effective radius of the galaxy and $R_d$ is 
the radius of the dark matter core) but massive 
($M_{dark}/M_{Lum}=10$) dark halo has been assumed for each galaxy. 
This particular configuration allows the development of a galactic wind in 
these small systems without destroying them.

The model allows one to follow in detail the evolution of the 
abundances of several chemical elements, starting from 
the matter reprocessed by the stars and restored into the 
interstellar medium (ISM) by stellar winds and type II and Ia supernova
explosions.

The main assumptions of the model are:

\begin{itemize}

\item
one zone with instantaneous and complete mixing of gas inside
this zone;

\item
no instantaneous recycling approximation, i.e. the stellar 
lifetimes are taken into account;

\item
the evolution of several chemical elements (H, D, He, C, N, O, 
Mg, Si, S, Ca, Fe, Ba and Eu) is followed in detail;

\item
the nucleosynthesis prescriptions include the yields of 
Nomoto et al. (1997) for type Ia supernovae, Woosley $\&$
Weaver (1995) (with the corrections suggested by 
Fran\c cois et al., 2004) for massive stars (M $> 10 M_{\odot}$), 
van den Hoek $\&$ Groenewegen (1997) for intermediate mass stars 
(IMS) and for Ba and Eu the ones described in LMC05 and Cescutti et al.
(2006).

\end{itemize}

The basic equations of chemical evolution are the same as described 
in LM03 and LM04 (see also Tinsley 1980, Matteucci 1996), as are 
the prescriptions for the SF (which follow a Schmidt law - Schmidt 1963),
initial mass function (IMF - Salpeter 1955), infall and galactic winds. 
The type Ia SN progenitors are assumed to be white dwarfs in binary 
systems according to the formalism originally developed by Greggio
$\&$ Renzini (1983) and Matteucci $\&$ Greggio (1986).
The main parameters adopted for the model of each galaxy, together with the 
predicted time for the occurrence of a galactic wind, $t_{GW}$, can be seen 
in Table 1.

\section{Results}

\subsection{The Sagittarius dSph galaxy} 

The Sagittarius dSph galaxy was characterized in LM03 and LM04 by
one long episode of SF (13 Gyr - as suggested by Dolphin, 2002) 
starting at the beginning of the formation of the system, 
and by intermediate values of SF efficiency ($\nu=1-5 \;Gyr^{-1}$) and 
very intense galactic winds, with efficiencies in the range $w_i$ = 9-13,
in order to reproduce several observed [$\alpha$/Fe] ratios
and the present total mass and gas mass.
With the aim of verifying this proposed evolutionary scenario,
we compare here the model as described in LM04,
with no modifications in the main parameters, with the most 
recent data concerning this galaxy. Monaco et al. (2005) 
claimed that the new [$\alpha$/Fe] data suggest a low star
formation rate (SFR) in Sagittarius, contrasting with
the values adopted for the SF efficiency in LM03. We will
show here that their first 
claim is true, namely that the low values of [$\alpha$/Fe] represents a 
low SFR,
but we will also show that the values for the SF efficiency, as suggested 
by LM03, 
do not contrast with that, due to the effects of the intense galactic 
wind on the star formation rate of the galaxy, which lowers 
substantially the SFR.

\begin{figure*}
\centering
\includegraphics[height=14cm,width=14cm]{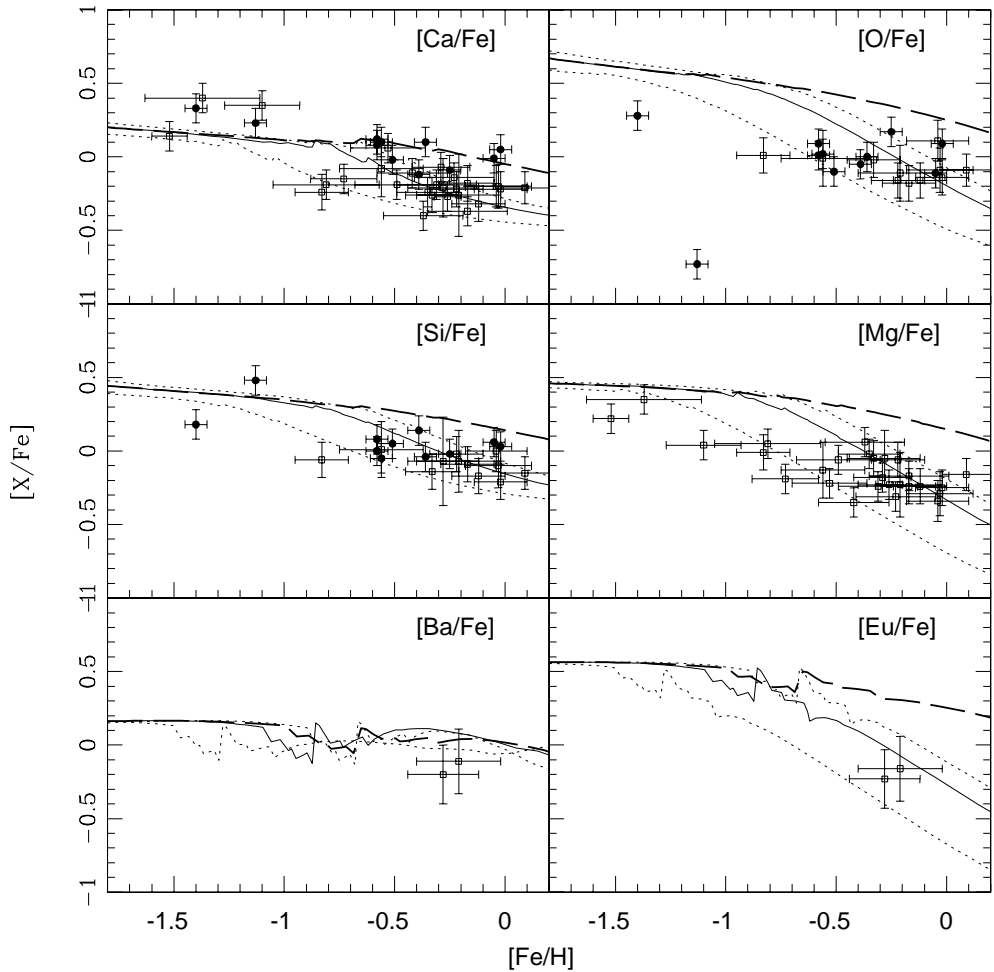}
\caption[]{[X/Fe] vs. [Fe/H] observed in Sagittarius dSph 
galaxy compared to the predictions of the chemical evolution
model for Sagittarius.The solid line represents the best model ($\nu =
3\;Gyr^{-1}$, w$_i$ = 9) and the dotted lines the lower ($\nu =
1\;Gyr^{-1}$) and upper  ($\nu = 5\;Gyr^{-1}$) limits for the SF
efficiency. The thick dashed line represents the best model without
galactic winds. The filled circles represent the data from Smecker-Hane
$\&$ Mc. William (1999) and the open squares the recent data from Bonifacio
et al. (2000, 2004) and Monaco et al. (2005).

} 
\end{figure*}

In Fig. 1, are shown the predictions of the models for 
Sagittarius with the same parameters as in LM04  
compared to the data of Bonifacio et al. (2000, 2004) and
Monaco et al. (2005) (open squares). We included also the 
data from Smecker-Hane $\&$ Mc. William (1999) (filled circles)
in order to get a better view of 
the difference between the two sets of data and how the model
predictions compare to them. As one can clearly see, the predictions 
of the models with galactic winds
for all four [$\alpha$/Fe] ratios analysed reproduce very well the 
observed data (old and recent), without any need of adjustment in the
main parameters of the models.

\begin{figure}
\centering
\includegraphics[height=8cm,width=8cm]{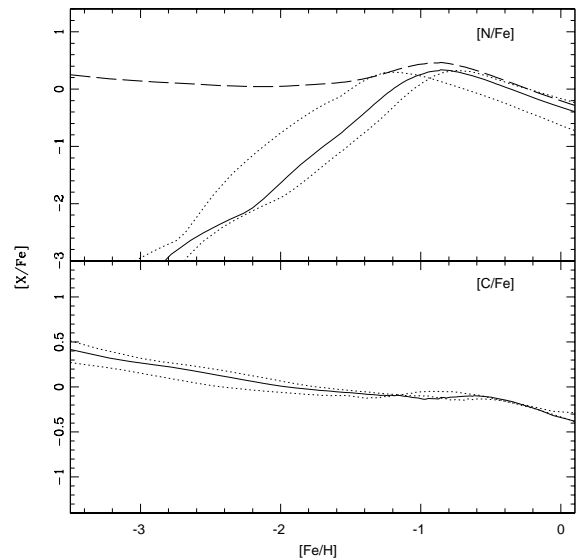}
\caption[]{[N/Fe] and [C/Fe] vs. [Fe/H] predicted by the chemical 
evolution
model for Sagittarius. The solid line represents the best model ($\nu =
3\;Gyr^{-1}$, w$_i$ = 9) and the dotted lines the lower ($\nu =
1\;Gyr^{-1}$) and upper  ($\nu = 5\;Gyr^{-1}$) limits for the SF
efficiency. The thick dashed line represents the best model with 
primary production of N in massive stars.} 
\end{figure}

As explained in the previous works (LM03, LM04), the behaviour of
all abundance ratios is controlled by the combined effects of the 
SF efficiency and
the galactic winds. These two features cannot be analysed separately,
due to the strong dependence between them. 
At the early
stages of the evolution of the galaxy, the massive stars, which die 
in the form of supernovae type II (SNe II) explosions, are the main 
contributors to the enrichment of the interstellar medium (ISM). This
contribution gives rise to high [$\alpha$/Fe] ratios since the main
products of SNe II explosion are the alpha-elements, whereas the elements
of the Fe-peak group are produced in a small amount in these explosions.
Besides that, with a high SF efficiency, the numbers of stars formed
is higher compared to the galaxies with low SFR and the galactic wind 
develops earlier. As soon as the efficient galactic
wind starts, the fraction of gas which fuels the SF decreases
intensively and, as a consequence, the SFR drops to low values.
With a much lower SFR, the number of new born stars is 
substantially lower and the injection of O in the ISM is almost 
ceased. The production and injection of 
Fe, on the other hand, continues, since its main production 
is the one occurring in supernova type Ia (SNe Ia)
explosions. The progenitors of these explosions (stars with masses
in the range $M= 1-8 M_{\odot}$) are characterized by long lifetimes, 
so they
continue exploding even after the decrease of the SFR. This scenario
results in a intense decrease in the [$\alpha$/Fe] ratios, as seen in
observed stars. The majority of these stars exhibit abundance ratios
which would place them in the [$\alpha$/Fe] diagram after the 
occurrence of the wind. In that case, the SFR when the stars would
have been formed is much lower (almost 5 times) than 
before the occurrence of the wind.

In that sense, the low values of [$\alpha$/Fe] observed in the majority 
of stars in Sagittarius do reflect a low SFR, but it does not mean that
the initial value of the SF efficiency should also be low. It should,
on the contray, be high, since the dSph galaxies are characterized by
very intense galactic winds, which removes a large fraction of the gas
reservoir, thus decreasing the SFR. Such a scenario, with 
a SF efficiency between $\nu=1-5\; Gyr^{-1}$ and a high wind efficiency
$w_i$ = 9-13, explains very well also the [Ba/Fe] and [Eu/Fe] ratios
observed in this galaxy. 

The observed patterns of both these ratios
are well explained also as the effects of the galactic
winds on the SFR and vice-versa and by the choice of the nucleosynthesis
prescriptions adopted. Europium is assumed to be produced only by
r-process (Woosley et al. 1994) in massive stars in the range 
$M= 10-30 M_{\odot}$, whereas Barium is assumed to be produced mainly
in low mass stars in the range $M= 1-3 M_{\odot}$ by s-process
(Busso et al. 2001) and a low fraction in
massive stars ($M= 10-30 M_{\odot}$) by r-process (see LMC05 
and Cescutti et al. 2006 for more details). At early stages of evolution
(low [Fe/H]), [Ba/Fe] and [Eu/Fe] exhibit values close to
or higher than solar due to the injection in the ISM
of Ba and Eu produced by r-process in massive stars (in the range
$M= 10-30 M_{\odot}$). Soon after the first supernovae type Ia 
(SNe Ia) start exploding, the abundance of Fe in the ISM increases,
the wind develops, the SFR decreases, the r-production of Ba 
and Eu is almost halted and, as a consequence, the [Ba/Fe] and [Eu/Fe] 
ratios suffer an abrupt decrease. The decrease in [Ba/Fe] is not 
so intense, though, since the s-production of Ba in low mass stars soon 
becomes important and continues even after the 
onset of the wind.

In order to compare different types of chemical evolution models
and to make clear the importance of
galactic winds on the evolution of this galaxy, we included also the 
predictions of a simple model with no galactic wind for all six 
abundance ratios analysed (the thick dashed line in Figure 1). At low
metallicities 
([Fe/H] $<$ -0.8) the behaviour of the models (with and without 
galactic wind) cannot be distinguished, but as the metallicity grows 
one can notice a meaningfull difference. While the models with winds 
exhibit a sharp decrease in the abundance ratios above [Fe/H] $\sim$
-0.8 dex, the model without wind is characterized by a smooth 
and constant decrease with no change in the slope. As mentioned above,
the sharp decrease is a consequence of the effects of the galactic wind
on the SFR and vice-versa. On the other hand, in the simple model, since 
there is no wind, the rate of decrease in the abundance ratios remains 
the same throughout the evolution of the galaxy and the lowest values of 
[$\alpha$/Fe] and [Eu/Fe] cannot be reproduced. This inadequacy of the 
simple model in reproducing the observed data makes clear the important
role played by the wind in the evolution of the dSph galaxies and
shows that it is very difficult to explain the lowest values of the abundance
ratios observed without invoking the occurrence of intense galactic winds.

\bigskip

{\bf Nitrogen and Carbon}

\bigskip\noindent

In Fig. 2, are shown the [N/Fe] and [C/Fe] ratios
predicted by the models of Sagittarius dSph as described 
previously. Although there is no observed data for
these ratios, it is interesting to see how they
behave in the proposed scenario for the evolution of the dSph 
galaxies, since there is a huge debate on the literature regarding
the production of these two elements. One of the most debated 
question concerns the origin of N.
The bulk of the secondary N comes from low and intermediate mass stars 
and the latter can produce also some primary N during the third 
dredge-up in 
conjunction with the hot bottom burning (Renzini $\&$ Voli, 1981).
On the other hand, massive stars should in principle 
produce a small fraction of N and all secondary. However, recent data for 
the solar vicinity as well as theoretical calculations suggest that  
N in massive stars should have a primary origin
(see Chiappini, Matteucci $\&$ Ballero 2005 and references therein). 
In this paper, we used two 
models with two different prescriptions for N production in
massive stars: one with only secondary N in massive
stars (thin line) and another with only primary N (thick dashed line), as suggested by Matteucci (1986),
which seems to reproduce very well the most recent 
metal-poor stars
of our Galaxy (Chiappini, Matteucci $\&$ Ballero 2005). 
In this scenario, the production
of N is fixed for all metallicities, with a yield per 
massive star of 0.065 $M_{\odot}$ of nitrogen. This prescription is 
rather ``ad hoc'' but it suggests what should be the right 
N production. Models 
of massive stars with rotation are promising candidates for such a production 
(Meynet \& Maeder, 2001).

The patterns of both [C/Fe] and [N/Fe] are similar to the ones
predicted by a model for the Milky Way in the solar vicinity
(Chiappini, Matteucci $\&$ Meynet 2003; Chiappini, Romano $\&$ Matteucci, 
2003): while [C/Fe] is almost constant, with a slight decrease, over 
the entire metallicity range, [N/Fe] instead  increases at low 
metallicities and reaches a sort of plateau at [Fe/H]
$\sim$ -1.0 dex when the production of N in massive stars is secondary, 
and remains
almost constant when a primary production of N in massive stars is assumed.
This similarity with the Milky Way is expected due the other
similarities seen in other abundance ratios and in the
predicted metallicity distribution of stars (LM04; LMC05). 
One can notice, however, a minor difference 
between the prediction of [N/Fe] in Sagittarius and in the solar 
vicinity: there is a slight decrease at high metallicities 
([Fe/H] $\sim$ -1.0 dex) in the Sagittarius predictions. This decrease
is a consequence of the effects of the galactic winds on the SFR,
similar to what is observed in [$\alpha$/Fe]. With the decrease of the
SFR, the formation of new stars is almost halted and,
after a certain time interval, also the production of N. The time interval
in this case is larger than for $\alpha$-elements, since N
is mainly produced in intermediate mass stars whereas the main
production sites of $\alpha$-elements are SNe II.
The derivation of C and N abundance in stars of Sagittarius
dSph would provide tight constrains to the evolution of this galaxy,
since almost all stars so far observed in Sagittarius would
be placed after the occurrence of the wind and would consequently
be characterized by this decrease in the [N/Fe] ratio. The controversy
regarding the primary or secondary 
production of N in massive stars, 
however, could be clarified only with observations of more 
metal-poor stars.

\subsection{Carina}

The chemical evolutionary history suggested for Carina dSph galaxy
seems more complicated than the one proposed for Sagittarius.
Investigations of the observed color-magnitude diagram of this galaxy
have revealed different stellar populations indicating that Carina
could have been characterized by more than one episode of SF 
(Smecker-Hane et al. 1996;
Hurley-Keller, Mateo, Nemec 1998; Hernandez, Gilmore, 
Valls-Gabaud 2000; Dolphin 2002; Rizzi et al. 2003). LM03 and 
LM04 adopted two long episodes (3 Gyr)
of star formation occurring at intermediate stages of the evolution of
the system (following Hernandez, Gilmore, Valls-Gabaud, 2000), with 
low efficiency ($\nu = 0.02 - 0.4\;Gyr^{-1}$) and very intense 
galactic winds with efficiencies from seven to eleven times 
larger than the SFR (w$_i$ = 7 - 11) (see Table 1 in LM04 for more details).
This scenario allowed the model to reproduce several [$\alpha$/Fe] 
ratios (LM04), the [Ba/Fe] and [Eu/Fe] ratios (LMC05), the present 
day total mass and gas mass, and also predict the stellar metallicity
distribution. 

The predicted stellar metallicity distribution was recently
compared to observations by Koch et al. (2005). These authors estimated
the metallicity (represented as [Fe/H]) of 437 stars through the Ca II 
triplet, a well know metallicity indicator. 
The comparisons between their observed distribution
and the one predicted by LM04 revealed a general good agreement, but also
some discrepancies related to the metal-rich and metal-poor tails
(see Fig. 13 on Koch et al. 2005). While the predicted LM04 
distribution succeeds in reproducing the initial steep rise toward higher 
metallicities and the prominent single peak, it predicts a larger number
of stars formed at metallicities lower than [Fe/H] $\sim$ -2.5 dex, 
and a too steep decrease in the number of stars formed at
metallicities higher than [Fe/H] $\sim$ -1.5 dex, thus underestimating the 
number of stars at the metal-rich tail of the distribution (see Fig. 3).

However, the calibration of
the Ca II lines and the transformation into Fe abundance contain several uncertainties
and need to be taken with caution. For example, Calcium and Iron 
are formed in totally different nucleosynthesis processes and, 
consequently, do not trace each other directly. The variations of 
[Ca/Fe] in the course of the evolution of the galaxy 
should be taken into account in the calibration. In the case
of Koch et al. (2005) distribution, this lack of knowledge leads to an 
uncertainty up to 0.2 dex. Besides that, the metallicities covered
in their distribution range from $\sim$ -3.0 dex to almost solar, but
their calibrating globular clusters cover only the range between 
$\sim$ -2.0 dex to $\sim$ -1.0 dex. Therefore, they extrapolate their
calibration in order to achieve the metallicities in the metal-poor 
and metal-rich tails of the distribution. A proper comparison between
model's predictions and the observed distribution should take these
uncertainties into account and one should keep them in mind whenever
comparing models predictions to the observations.

By taking these facts into consideration and assuming that Koch's 
stellar metallicity distribution is the most accurate one available
nowadays,
one could argue then that these discrepancies could be related to 
the choices of some of the parameters in the LM04 models, particularly 
the SFH and the galactic wind efficiency, as we will see in the following. 

\begin{figure}
\centering
\includegraphics[height=8cm,width=8cm]{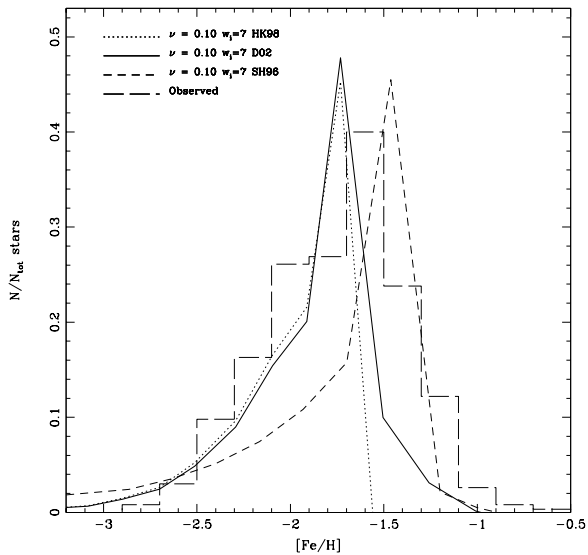}
\caption[]{The predicted metallicity distribution of stars by 
the models for Carina dSph with the LM04 parameters 
($\nu = 0.1\;Gyr^{-1}$ and w$_i$ = 7), but with 
different SFH: Hulley-Keller et al. (1998)(dotted line), Smecker-Hane 
et al. (1996)(dashed line) and Dolphin (2002)
(thick solid line), compared to the observed data (long dashed line).  }
\end{figure}

The galactic wind efficiency adopted in the best model of LM04 is 
seven times the SFR, chosen to reproduce the observed decrease in
several [$\alpha$/Fe] ratios and the estimated present day gas mass.
This high wind efficiency leads also to the abrupt decrease in the 
metallicity distribution at high [Fe/H], consequently a smaller value 
might reduce the intensity of the decrease and still reproduces the 
gas mass and the abundance ratios. Besides that, LM04 adopted the SFH
inferred by Hernandez, Gilmore and Valls-Gabaud (2000), which suggests 
a SF in two long episodes (with 3 Gyr duration) occurring at galactic
ages of 6 and 10 Gyr. Other authors, on the other hand, suggest different
scenarios for the evolution of Carina. While Smecker-Hane et al. (1996)
and Hulley-Keller, Mateo \& Nemec (1998) infer three episodes of SF, 
occurring respectively at 2-4, 9-12 and 13 Gyr, and 0-1, 7-9 and 12 Gyr,
Dolphin (2002) claims that the SF in Carina proceeded in only one very long
episode. The different SFHs can also affect the stellar metallicity 
distribution but one should be aware of the uncertainties in the
derived SFHs. 

The methods applied in the SFH determinations are based on
the analysis of colour-magnitude diagrams of the resolved stellar 
populations of the stars in the galaxies. However, a certain number of 
assumptions is 
necessary in this procedure, such as initial mass function and metallicity.
The results also rely strongly on the set of isochrones which are 
adopted. 
Besides that, there are several hints that the stellar population in 
dSph and its properties (for example, metallicities and abundance ratios) 
might vary with radius in a galaxy (Harbeck et 
al. 2001; Tolstoy et al. 2004). In that sense, observations of different 
small regions of the same galaxy might lead to a different 
color-magnitude diagram and, consequently, to a different SFH. 
Ideally, in order to get a more accurate result, one
should cover the entire galaxy, or, at least, a large fraction of the field 
(see Rizzi et al. 2003), but that is not always possible.
Such difficulties lead to several uncertainties, which are reflected in
the different SFHs determined from different groups for the same galaxy, 
as in the case of Carina.
One way of distinguishing the most reliable 
SFH is to make use of a chemical evolution model with the adopted SFHs 
and compare the predictions of such models with well established
observational constraints. Here we intend to show that an accurate
stellar metallicity distribution could be a very useful tool in the
attempt to better constrain the SFHs of the dSph galaxies.

Another parameter which affects
the predicted metallicity distribution is the IMF. In LM04, we
used a Salpeter IMF, but there are suggestions (Aloisi, Tosi, Greggio, 
1999; Chiappini, Romano, Matteucci 2003) that dwarf galaxies 
might be well represented by IMF with a flatter slope, for instance x=1.1.

\begin{figure}
\centering
\includegraphics[height=8cm,width=8cm]{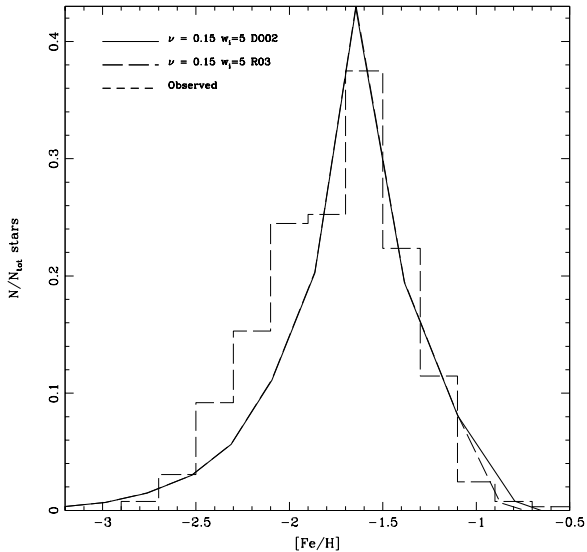}
\caption[]{The metallicity distribution of stars predicted by
the models for Carina dSph with $\nu = 0.15\;Gyr^{-1}$,
 w$_i$ = 5, and the SFH of Dolphin (2002) (thick solid line)
and Rizzi et al. (2003) (long dashed line),  
compared to the observed data (dashed line).  }
\end{figure}

Here we test the effects of changing these parameters on the shape of 
the stellar metallicity distribution.

As a first step, we test a new model with the same values for the SF and 
wind efficiencies, the same SFH, but with a Salpeter like IMF with
slope x=1.1. In such a case, the predicted 
metallicity distribution exhibits a decline at high [Fe/H] ($>$-1.5 dex) 
less intense than in LM04, in better agreement with  the distribution of 
Koch 
et al. (2005). This model, however, continues to predict a number of
metal-poor stars higher than observed (the so called G-dwarf problem) and,
besides that, predicts too high [O/Fe] and [Mg/Fe] ratios when compared
to the observed ones. This is a consequence of the higher number of 
massive stars formed when a IMF with this slope is adopted. We suggest,
therefore, that the Salpeter IMF is still the best choice to reproduce
the observational data of Carina dSph galaxy.

We then run  models with classical Salpeter IMF but adopting different SFHs, 
in particular those from
Smecker-Hane et al. (1996), Hulley-Keller, Mateo, Nemec (1998),
and Dolphin (2002), without changing the values of the other
parameters of the LM04 best model (namely w$_i$ = 7, $\nu$ = 0.1 
Gyr$^{-1}$). One can see in Fig. 3 that models with different SFHs
predict similar metallicity distributions, 
but with differences which are important enough to discriminate between
different  
scenarios when confronted to the observed metallicity distribution.
The models with SF beginning as soon as the gas collapses to form the 
galaxy (Dolphin 2002, and Hulley-Keller et al.
1998), do not exhibit the so called G-dwarf 
problem, i.e. the number of metal-poor stars predicted is similar 
to what is observed. When, on the other hand, the SF begins some Gyr
after the gas has started collapsing (Smecker-Hane et al. 1996 SFH), 
there
is an overproduction of the number of stars with low [Fe/H]. However, 
even though the SFH of Dolphin (2002) and Hulley-Keller et al.
(1998) solve the problem at the metal-poor tail of the metallicity
distribution, they are not able to reproduce the metal-rich tail. 

The model with the SFH from Hulley-Keller 
et al. (1998) predicts a very intense decline, even more intense than in
LM04, whereas the model with Dolphin (2002) SFH predicts a smoother 
decline in better 
agreement with observations. 

In Fig. 4, the observed metallicity distribution is compared to a model
with the Dolphin (2002) SFH and a lower galactic wind efficiency
(w$_i$ = 5 instead of 7) coupled with a SF efficiency marginally higher
($\nu$ = 0.15 instead of 0.10 Gyr$^{-1}$). The SF efficiency was increased
in order to get a better fit to the observed distribution. With this new
model the agreement between observations and model's prediction is 
very good. Not only the peak of the distribution is very well 
reproduced but also
the metal-rich and metal-poor tails are fitted. As expected, the
"G-dwarf problem" vanished due to the new SFH and the decline of the 
metallicity distribution is smoothed thanks to a combination of a lower
galactic wind efficiency and a long initial star formation period.
The adopted SFH, in fact, is almost continuous, even though there are 
several 
observational hints indicating that Carina must have undergone
separate episodes of SF. 

Given this fact and the limitation of the Dolphin SFH, due to the 
small
field of the galaxy covered by his observations, we adopted also the 
SFH of Rizzi et al. 2003. These authors, based on wide-field observations,
suggested a SFH which is characterized by several episodes of star 
formation. Actually they proposed three similar SFHs depending on the 
adopted choice for the evolution of metallicity. The models 
with the three SFH provide very similar abundance ratios and 
metallicity distributions. The model with the SFH with constant 
metallicity though is the one with the best adjust to the observational
data, so we show from now on the results of this model.
In Figure 4, we show the stellar metallicity 
distribution of a model with Rizzi's SFH
(long dashed line). 
As one can see, it is almost identical to the one with 
Dolphin's SFH, with a minor difference at the high-metallicity tail. 
The similarity between the two distributions is a result of the fact that
the majority of stars are formed in the first Gyr of the evolution of the 
galaxy.
Since Dolphin's SFH is almost constant and Rizzi's one has two initial 
long episodes of activity ($\sim$ 2 Gyr each), they form a similar number 
of stars in the early evolution of the system ([Fe/H] $<$ -1.2 dex).
After these two initial episodes, the interval with no activity 
in Rizzi's SFH would generate a difference in the predictions 
(compared to a continuous SF) only at higher 
metallicities ([Fe/H] $>$-1.2 dex), as seen in the high-metallicity
tail of the stellar metallicity distribution. 
After a few Gyr of activity, the galactic wind develops and 
removes a large fraction of the gas from the galaxy in such a way that
the subsequent SF activity is not strong enough to form a significant
number of metal-rich stars. Since Rizzi's SFH is based on wide-field 
observations covering a large fraction of the galaxy and predicts a 
scenario more similar to what is commonly adopted for 
Carina, we adopted this choice of SFH instead of the small field
SFH of Dolphin, which actually does not seem to be realistic (see 
the recent SFH proposed for Carina in Dolphin et al. 2005).

The model with Rizzi's SFH and the new values for the wind and 
SF efficiencies
should also be able to reproduce the observed abundance ratios
in order to be successful. With this aim, we show in
Fig. 5  the predictions of the new Carina model as described above
for [Eu/Fe], [Ba/Fe] and several [$\alpha$/Fe], compared to the 
observations. One can clearly see that this model
still reproduces very well all abundance ratios analysed here. 
The fit is not unexpected since 
the adopted variations of $\nu$ and $w_i$ are well inside 
the ranges suggested for these parameters by LM04.

\begin{figure*}
\centering
\includegraphics[height=14cm,width=14cm]{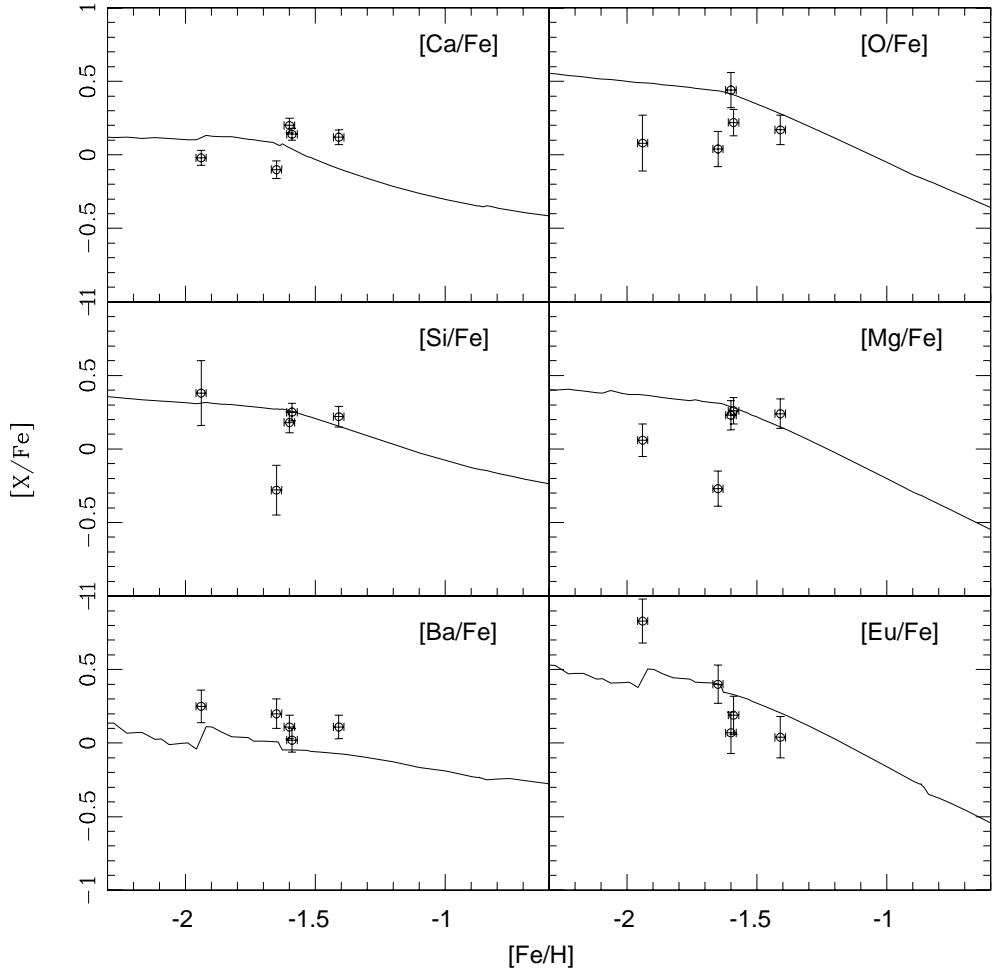}
\caption[]{[X/Fe] vs. [Fe/H] observed in Carina dSph 
galaxy compared to the predictions of the new
model for Carina ($\nu = 0.15\;Gyr^{-1}$, w$_i$ = 5) with 
the SFH of Rizzi et al. (2003).}
\end{figure*}

\bigskip
{\bf Nitrogen and Carbon}

\bigskip\noindent

After showing the good agreement with the observations obtained
by means of the predictions 
of the Carina model for several abundance ratios and for the
metallicity distribution of stars, we use the same model to predict
the evolution of [N/Fe] and [C/Fe] as a function of [Fe/H] in Carina
dSph galaxy. These predictions 
are shown in Fig. 6. As in the case of Sagittarius
two scenarios for the production of N in massive stars
(primary and secondary -
thick dashed and thin solid lines, respectively) were adopted.
The patterns of both ratios are in general similar to the ones
predicted for Sagittarius: while [C/Fe] is almost constant
in the entire metallicity range, [N/Fe] increases at low
metallicities, reaches a peak at [Fe/H] $\sim$ -1.8 dex, and 
then decreases. There are, however, significant differences 
between the predictions for Carina and Sagittarius, especially
in the case of [N/Fe]: the increase happens at lower 
metallicities ([Fe/H] $\sim$ -3.4 dex) in the case of Carina, 
the peak of the ratio is reached faster and the decrease,
which is more steep, occurs also at lower metallicities. 
These differences are related to the 
values of the SF efficiencies adopted for both galaxies:
the Carina model is characterized by a low SF efficiency
($\nu$ = 0.15 $Gyr^{-1}$), whereas in Sagittarius best model the 
adopted value is much higher ($\nu$ = 3.0 $Gyr^{-1}$).
In the case of Carina, the derivation of the abundance of N
in the observed stars would also not help solving the problem 
regarding its production, since these stars exhibit metallicities
close to [Fe/H] $\sim$ -2.0 dex. At this metallicity range, 
the models for primary and secondary N predict similar
[N/Fe]. Only observations of stars at lower metallicities
would help in the attempt to disentangle these two productions
of N.

\begin{figure}
\centering
\includegraphics[height=8cm,width=8cm]{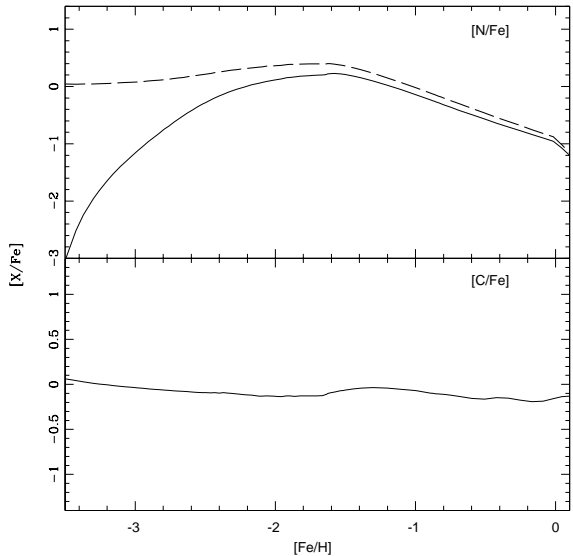}
\caption[]{[N/Fe] and [C/Fe] vs. [Fe/H] predicted by the chemical evolution
model for Carina. The thin solid line represents the new model 
($\nu = 0.15\;Gyr^{-1}$, w$_i$ = 5) with secondary production of N and 
the thick dashed line represents the same model with primary production
of N in massive stars.}
\end{figure}

\subsection {The [N/O] ratio}

Due to the different time-scales for the production of 
N and O, [N/O] is one of the most used abundance 
ratio in the investigation of the nature of N and of the 
chemical evolution of several types of galaxies. In order 
to get a better picture of the evolution of the dSphs 
studied here and to make an easier comparison with other 
types of galaxy and among the dSph themselves, we 
predict the evolution of [N/O] as a function of [O/H]
for the models of Carina (lower panel in Fig. 7)
and Sagittarius (upper panel in Fig. 7) dSph galaxies.
As in the case of [N/Fe],
two scenarios for the production of N were adopted. 
As expected, the predicted [N/O] increases as a function of 
increasing [O/H], when primary N from massive stars is assumed. 
At early stages of evolution, the injection of O
in the ISM is predominant, due to the short time-scale of the progenitors
of SNe II (the main production site of O), thus giving rise to low
[N/O] ratios. As the evolution proceeds, the production of secondary N
in massive stars increases, then primary plus secondary 
N is released in to the ISM 
also by intermediate mass stars and, as a consequence,
the values of [N/O] no longer increase. If one considers,
on the other hand, the production of primary N in 
massive stars, then [N/O] remains high over the entire 
evolution of the system. After the development of the 
galactic winds, the [N/O] values decrease (specially in 
the case of Carina) due to the effects of the wind on 
the SFR, as in the case of [N/Fe]. This decrease
is clearly seen in the predictions of Carina's 
model, but hardly seen in Sagittarius. Once more, this
difference, and the [N/O] early increase in Carina, 
are related to the values adopted for the SF efficiencies 
in both galaxies. In both cases, though, the pattern 
of the [N/O] ratio is similar: it increases until
a peak is reached, and then a plateau or a slight decrease 
appears. A quite different behavior for N/O has been suggested by
by Kawata et al. (2005), who predicted
for Draco dSph a very high [N/O] (up to 2.0 dex) at low 
metallicities ([Fe/H] $\sim$ -3.5 dex) followed by a constant 
decrease of [N/O] with increasing oxygen abundance. Such a
behaviour is very hard to explain in the light of the known 
nucleosynthesis, especially those very high 
values of [N/O] at very low metallicities. The authors claim a 
differential wind which carries away more oxygen that nitrogen but 
also this behaviour is difficult to understand, since it would imply an 
inversion in the oxygen abundance, not seen in their plot.

\begin{figure}
\centering
\includegraphics[height=8cm,width=8cm]{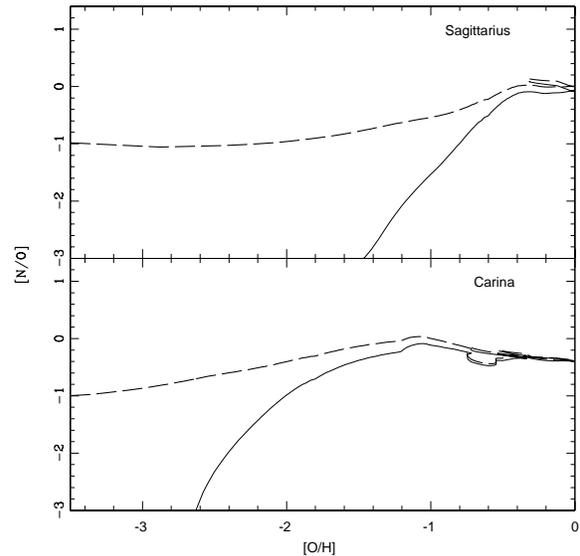}
\caption[]{[N/O] vs. [O/H] predicted by the models of
Carina (lower panel) and Sagittarius (upper panel), 
with primary (thick dashed line) and secondary production
(thin solid line) of N.}
\end{figure}

\section{Summary}

In order to better test the scenario for the chemical evolution history
of Carina and Sagittarius dSph galaxies proposed by LM03 and LM04
we compared the predictions of these models with the most recent data
concerning these two systems. In particular, we compared the predictions
of the Carina model with several observed abundance ratios and with 
the stellar metallicity distribution recently published by Koch et al. 
(2005).
In the case of Sagittarius, the model predictions were compared 
to the previous data from Smecker-Hane $\&$ Mc. William (1999)
and  to the recent data from Bonifacio et al. (2000, 2004)
and Monaco et al. (2005). In both cases the models reproduce very
well the new data, without any need of modification in the case of
Sagittarius and with minor changes in the SF and galactic wind 
efficiencies
of the LM04 best model for Carina. In Carina's model we also adopted a 
different SFH, namely the one of 
Rizzi et al. (2003), which results in a 
very good agreement with the observed metallicity distribution of stars.
In the two models the proposed scenarios for the
evolution of these galaxies are very similar: they form by means of a
continuous infall of primordial gas until a critical mass is reached; the 
SF is characterized by a long episode of activity in the case
of Sagittarius and several episodes in the case of Carina, with low to
intermediate efficiencies;
the galactic winds are very efficient (5 times the SFR in Carina
and from 9 to 13 times the SFR in Sagittarius) and crucial in the
evolution of these galaxies. The effects of the winds on the SFR and
vice-versa are the main processes which define the abundance ratio 
patterns and the shape and peak of the metallicity distribution 
of stars.

The choice of the best range of values for the SF and wind 
efficiencies is made in order to get the best fit to the observational 
constraints of each galaxy. Obviously,
it is not expected that all galaxies be characterized by the same values
for these two parameters, since different galaxies would have had 
different initial conditions. 
However, the similarity of the predicted (and observed) [$\alpha$/Fe] vs. 
[Fe/H] relations in these galaxies suggests that the details of the star
formation and wind histories are not important: what matters are the 
average star formation and wind rates over the galactic lifetime. In 
fact, abundances do not depend directly upon the rate of star formation 
and rates of gas flows (in and out), which in turn influence the SFR,
but rather upon the integral of the star formation rate.

The main conclusions can be summarized as follows: 

\begin{itemize}

\item
the Sagittarius chemical evolution model proposed by LM04 is able
to reproduce both the previous available data and the most recent 
data concerning several [$\alpha$/Fe]
ratios without any modification in the main parameters:
$\nu$ = 1.0-5.0 $Gyr^{-1}$, 
$w_{i}$ = 9-13. Even though the values suggested for the SF 
efficiency are not low, when the winds starts the SFR decreases
substantially (almost 5 times) due to the fast removal of a large fraction 
of the gas content of the galaxy. Since most of the observed stars 
would have been formed just after the beginning of the wind, their 
low values of [$\alpha$/Fe] reflect this decrease in the SFR;

\item
the chemical evolution model for Carina dSph is able to reproduce
very well the observed metallicity distribution and abundance ratios
if some small changes are made relative to the LM04 model. 
A different SFH (the one 
from Rizzi et al. (2003), 
a lower galactic wind efficiency (w$_i$ = 5) and a marginally 
higher SF efficiency ($\nu$ = 0.15 Gyr $^{-1}$)
enables us to predict a number of stars at the metal-poor tail of the 
distribution similar to what is observed, solving then the so called 
G-dwarf problem, one distinct peak at the same location as the observed
one and
a smooth decline at the metal-rich tail. Several observed abundance 
ratios (such as [Ba/Fe], [Eu/Fe] and several [$\alpha$/Fe]) 
are also reproduced with this choice of parameters;

\item
in the proposed scenario for the evolution of Carina, as in the case of
Sagittarius (and other dSphs), the
effects of the galatic winds on the SFR and vice-versa are crucial
in determining the evolution of the abundance ratios and affect substantially
the shape, peak and slope of the stellar metallicity distribution at 
the high-metallicity end.

\item
we also show new predictions for [N/Fe], [N/O] and [C/Fe] ratios for the 
two galaxies. In the case of Sagittarius there is a similarity 
with the predictions for the solar neighbourhood,
but with a slight decrease of [N/Fe] at high metallicities due 
to the effect of the galactic wind on the star formation rate, 
effect not present in the Milky Way. In Carina, the patterns 
of these ratios are similar to the ones of Sagittarius, except for 
larger values of [N/Fe] at low metallicities and a more visible 
decrease of [N/Fe] and [N/O] at high metallicities. The 
differences between the predictions for these two galaxies 
are mainly a consequence of the different SF efficiencies adopted for 
each galaxy. 

\end{itemize}

\section*{Acknowledgments}
G.A.L. acknowledges financial support from the Brazilian agency 
FAPESP (proj. 04/07282-2). F.M. acknowledges financial support  
from COFIN2003 from the Italian 
Ministry for Scientific Reasearch (MIUR) project ``Chemical 
Evolution of Galaxies: interpretation 
of abundances in galaxies and in high-redshift objects''.

\end{document}